\documentclass[twocolumn]{aastex63}
\usepackage{gensymb}
\usepackage{hyperref}

\accepted{November 16, 2021}

\shorttitle{Sodium Chloride on Europa from UV Spectroscopy}
\shortauthors{Trumbo et al.}

\graphicspath{{./}{figures/}}

\begin{document}

\title{A New UV Spectral Feature on Europa: Confirmation of NaCl in Leading-hemisphere Chaos Terrain}

\correspondingauthor{Samantha K. Trumbo}
\email{skt39@cornell.edu}

\author[0000-0002-0767-8901]{Samantha K. Trumbo}
\affiliation{Division of Geological and Planetary Sciences, California Institute of Technology, Pasadena, CA 91125, USA}
\affiliation{Cornell Center for Astrophysics and Planetary Science, Cornell University, Ithaca, NY 14853, USA}

\author[0000-0002-1559-5954]{Tracy M. Becker}
\affiliation{Southwest Research Institute, 6220 Culebra Road, San Antonio, TX 78238, USA}

\author[0000-0002-8255-0545]{Michael E. Brown}
\affiliation{Division of Geological and Planetary Sciences, California Institute of Technology, Pasadena, CA 91125, USA}

\author[0000-0003-4752-0073]{William T. P. Denman}
\affiliation{Division of Geological and Planetary Sciences, California Institute of Technology, Pasadena, CA 91125, USA}

\author[0000-0002-4725-4775]{Philippa Molyneux}
\affiliation{Southwest Research Institute, 6220 Culebra Road, San Antonio, TX 78238, USA}

\author[0000-0002-0435-8224]{Amanda Hendrix}
\affiliation{Planetary Science Institute, Tucson, AZ 85719, USA}

\author[0000-0001-9470-150X]{Kurt D. Retherford}
\affiliation{Southwest Research Institute, 6220 Culebra Road, San Antonio, TX 78238, USA}

\author[0000-0003-0554-4691]{Lorenz Roth}
\affiliation{Space and Plasma Physics, KTH Royal Institute of Technology, Stockholm, Sweden}

\author[0000-0003-1459-3444]{Juan Alday}
\affiliation{AOPP, Department of Physics, University of Oxford, Oxford, UK}

\begin{abstract}

Recent visible-wavelength observations of Europa's surface obtained with the Hubble Space Telescope revealed the presence of an absorption feature near 450 nm that appears spatially correlated with leading-hemisphere chaos terrain. This feature was interpreted to reflect the presence of irradiated sodium chloride ultimately sourced from the interior. Here, we use ultraviolet spectra also collected with the Hubble Space Telescope to detect an additional previously unseen absorption near 230 nm, which spatially correlates with the 450 nm feature and with the same leading-hemisphere chaos terrain. We find that the new ultraviolet feature is also well-matched by irradiated sodium chloride at Europa-like conditions. Such confirmation of sodium chloride within geologically young regions has important implications for Europa's subsurface composition.
\end{abstract}

\keywords{Galilean satellites (627), Europa (2189), Planetary surfaces (2113), Surface composition (2115)}

\section{Introduction} \label{sec:intro}
Salt hydrates ultimately sourced from Europa's putative subsurface ocean have long been proposed as components of its geologically young surface \citep[e.g.][]{McCord1998_science}, which could allow remote-sensing studies to constrain the ocean's composition and chemical history. Observations of Europa's tenuous atmosphere revealed the presence of sodium and potassium \citep{Brown1996, Brown2001}, and subsequent observations and modeling have suggested a potential link to surface salts and, ultimately, the ocean below \citep[e.g.][]{Johnson2000,Leblanc2002,Johnson2002,Leblanc2005}. However, the precise identification of Europa's endogenous surface salts has been an ongoing subject of debate. Numerous studies interpret \textit{Galileo} Near-infrared Mapping Spectrometer (NIMS) spectra to reflect endogenous sulfates \citep[e.g.][]{McCord1998_science, Dalton2005, Dalton2007, Dalton2012}, while more recent, ground-based spectroscopic studies have either favored a chloride-dominated interpretation \citep{BrownHand2013, Fischer2015, Fischer2016} or, in the interpretation of \citet{Ligier2016}, a mixture of magnesium chloride, chlorate, and perchlorate. 

The persistence of this debate has been facilitated, in part, by the paucity of compositionally diagnostic absorptions in 1--2.5 $\micron$ spectra of Europa's surface. Discrete features of many of the proposed sulfates were below the spectral resolution of NIMS or could be smoothed out with sufficiently complex linear-mixture modeling \citep[e.g.][]{Dalton2012}, while similar modeling approaches, rather than characteristic absorptions, were used to infer the compositions of \citet{Ligier2016}. In contrast, like Europa, the proposed sodium and potassium chloride of \citet{BrownHand2013} are simply spectrally smooth in the near infrared, which makes them hard to identify in such data, even when present in high abundance.

Though smooth in the infrared, alkali chlorides do form spectrally distinct ``color-center" absorptions at visible-wavelengths when subjected to energetic particle irradiation \citep{HandCarlson2015, Poston2017, Hibbitts2019}, a process that occurs across the entire surface of Europa \citep{Paranicas2009}. Indeed, recent spatially resolved visible-wavelength spectra from the Hubble Space Telescope (HST) detected a previously unseen absorption feature near 450 nm that is consistent with the F-center of irradiated sodium chloride (NaCl) \citep{TrumboEtAl2019}. The NaCl F-center feature appears only on the leading hemisphere, sheltered from the magnetospheric sulfur implantation and radiolytic alteration of the trailing hemisphere \citep{Carlson2002, Paranicas2009}, and corresponds particularly to Tara Regio, a large, visibly discolored region of geologically young ``chaos" terrain \citep{Doggett2009, Leonard2018}. This distribution strongly suggests an interior origin, implying that the F-centers are forming in relatively pristine endogenous material containing NaCl.

Here, we use mid-UV HST spectra of Europa's surface to detect a previously unseen absorption near 230 nm, which geographically correlates with the recently detected F-center feature. The parallel laboratory work of \citet{Brown2022} shows that NaCl irradiated under Europa-like conditions forms an additional feature, known as the V$_3$ center, at similar wavelengths. We find that the V$_3$ band matches our observed feature well and argue that this detection confirms the presence of NaCl within geologically young chaos regions on Europa. The nondetection of the 230 nm feature in previous mid-UV datasets can be explained by their lack of high-quality, spatially resolved coverage of Europa's leading hemisphere at these wavelengths. The International Ultraviolet Explorer (IUE) had poor signal-to-noise below $\sim$240 nm \citep{Nelson1987, Domingue1998} and the \textit{Galileo} UVS observations of the leading hemisphere were either disk-integrated or challenged by the radiation environment at Europa \citep{Hendrix1998, Carlson1999Perox}.  A complete analysis of the full mid-UV HST dataset, which presents the first high-quality, spatially resolved coverage of Europa in the 200--240 nm range, is given in \citet{Becker2022}.

\section{Observations and Data Reduction} \label{sec:observations}
We present HST observations of Europa's surface made with the Space Telescope Imaging Spectrograph (STIS) across four HST visits. Table \ref{table:obs} lists the corresponding dates, times, and geometries of each visit. The 52$^{\prime\prime}$ x 0.2$^{\prime\prime}$ slit was used in two pointings per visit in order to cover the central $\sim$40$\%$ of the disk. The G230L low-resolution, first-order grating (R$\sim$500; spectral range: 157--318 nm) was used with exposure times of 987.2 seconds per slit position for the 2018 visits and 777.5 or 808.2 seconds for the 2019 visit. This approach provided high-quality, spatially resolved spectra from 200--300 nm, with shorter wavelengths ($<$200 nm) dominated by noise.

In our analysis, we used the flux- and wavelength-calibrated data provided by HST after standard reduction with the STIS calibration pipeline (calstis). We extracted single spectra by taking individual rows from the two-dimensional spectral images, corresponding to the 0.025$^{\prime\prime}$ pixel-scale (equal to the 0.025$^{\prime\prime}$ or $\sim$80-km diffraction-limited resolution at 230 nm). We then divided each by a corresponding solar spectrum from the same date of observation, as obtained by the SORCE/SOLSTICE ultraviolet spectrometer \citep{McClintock2005} and downloaded from the LASP Interactive Solar Irradiance Datacenter (LISIRD). Prior to division, we smoothed and binned the solar spectra to match the STIS spectral resolution and sampling. Finally, we calculated the corresponding latitude/longitude coordinates of each extracted pixel using the aperture geometry information in the HST FITS headers and the phase, angular size, and pole orientation of Europa from JPL Horizons for the time of each exposure.

\begin{table}
\begin{center}
\caption{Table of Observations\label{table:obs}}
\begin{tabular}{ccccc}
\hline\\[-4mm] \hline
Date&Time&Central&Central&Angular\\
(UT)&(Start/End)&Lon.&Lat.&Diameter\\ \hline
2018 Apr 11 & 07:09/07:29 &  197$\degree$W & -3.89$\degree$ & 0.95$^{\prime\prime}$\\
2018 Apr 19 & 02:42/03:02 &  270$\degree$W & -3.88$\degree$ & 0.96$^{\prime\prime}$\\
2018 Jun 1 & 10:03/10:23 &  345$\degree$W & -3.74$\degree$ & 0.96$^{\prime\prime}$\\
2019 Jun 21 & 07:23/07:42 &  91$\degree$W & -3.14$\degree$ & 1.00$^{\prime\prime}$\\ \hline
\end{tabular}
\end{center}
\end{table}

\section{Feature Identification and Mapping} \label{sec:mapping}
\begin{figure}[t]
\figurenum{1}
\plotone{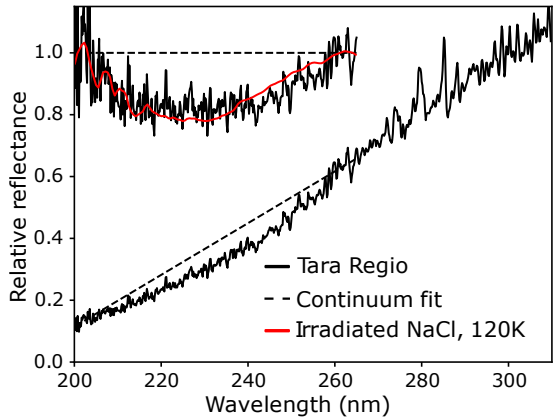}
\caption{Average HST/STIS spectrum of Tara Regio showing the 230 nm feature, which is consistent with the V$_3$ center of  irradiated NaCl. The dashed line is a linear continuum fit, and the continuum-removed feature is included above the spectrum. The continuum-removed NaCl V$_3$ band from the 120 K laboratory spectrum of \citet{Brown2022} is overlain in red and scaled to match the depth of the observed feature in the HST data. \label{fig:spec}}
\end{figure}

The HST spectra of some locations on Europa reveal a subtle absorption near 230 nm (Figure \ref{fig:spec}), which appears discrete against the sloping continuum in the mid-UV and at roughly the wavelength of the NaCl V$_3$ band at 120 K \citep{Brown2022}. To facilitate comparison with the laboratory data, we remove linear continua from both an average HST spectrum containing the feature (Figure \ref{fig:spec}) and a 120 K laboratory NaCl spectrum from \citet{Brown2022} using identical bounds. We find that a linear continuum works best for both the HST and the laboratory data and seems most reasonable based on leading-hemisphere spectra that do not contain the 230 nm absorption. We thus fit a linear continuum from 200 to 265 nm, excluding the 202--260 nm region, which corresponds to the feature in both spectra, from the fit. As the laboratory data do not extend below 200 nm and the HST spectra are of very poor quality at those wavelengths, it appears likely that these bounds do not entirely capture the extent of the feature on the short-wavelength end. The result of the laboratory comparison is included in Figure \ref{fig:spec}, which demonstrates that the band minimum, shape, and apparent width of the observed 230 nm feature are well-matched by the V$_3$ center of laboratory-irradiated NaCl at 120K. Indeed, we obtain equivalent, marginally different fits using any of the spectra of \citet{Brown2022}, which show slight variations in the V$_3$ band with different laboratory conditions.

Though the 230 nm absorption spectrally matches NaCl quite well, its geographic distribution in relation to the proposed NaCl F-center is also critical to cementing this identification. In order to map the 230 nm feature, we measure its integrated band area in each HST spectrum after fitting and removing a linear continuum as described above. As we are now dealing with individual pixels rather than a high signal-to-noise average, we slightly adjust the bounds for continuum removal to 200--285 nm, excluding the 215--275 nm region across the absorption, in order to minimize the effects of increased noise at the short-wavelength end of the spectra. We then integrate the residual feature to obtain the band area or equivalent width (i.e. the width
of a 100\% depth absorption of the same integrated area) for each pixel, and map the resultant values across the surface of Europa using the coordinates as obtained in Section \ref{sec:observations}. 

\begin{figure}[t]
\figurenum{2}
\plotone{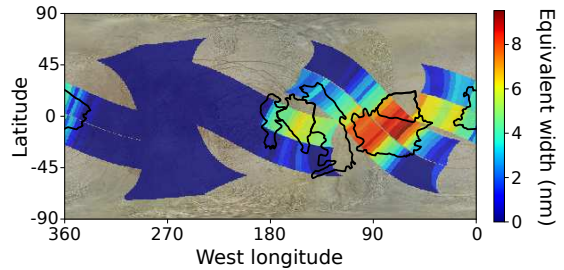}
\caption{Map of Europa's 230 nm feature. This newly detected absorption appears exclusively on the leading hemisphere. Black outlines indicate large-scale leading-hemisphere chaos regions, mapped approximately from \citet{Doggett2009}. The strongest 230 nm absorptions are associated with the chaos region Tara Regio ($\sim$85$\degree$W), which also contains the strongest NaCl F-center absorptions seen in previous visible-wavelength HST spectra \citep{TrumboEtAl2019}. Background image credit: NASA/JPL/Björn Jónsson \label{fig:map}}
\end{figure}

We find that the largest 230 nm absorptions are nearly perfectly constrained to Tara Regio (Figure \ref{fig:map}), the same leading-hemisphere chaos region in which the proposed NaCl F-center absorption is strongest \citep{TrumboEtAl2019}. Similar to what was seen for the F-center, we also find weaker absorptions across the other large-scale leading-hemisphere chaos west of Tara Regio, as well as immediately outside of Tara Regio, which presents a mixture of smaller-scale chaos and lenticulated terrain amongst older plains \citep{Doggett2009, Leonard2018}. As discussed in \citet{TrumboEtAl2019}, color-centers indicate irradiated NaCl, so their distribution on Europa should reflect a convolution of NaCl abundance and radiation intensity. Chlorides emplaced onto the leading hemisphere would be irradiated by $\geq$20 MeV electrons, which impact in an equatorial lens centered on the leading apex \citep{Nordheim2018}, providing the energy for color-center growth. Such a convolution may explain why Tara Regio, in particular, exhibits the strongest features, as well as why comparable absorptions to those observed in other leading chaos can be seen in more mixed low-latitude terrain. The concurrent match to NaCl in both wavelength and geography provides strong evidence that the 230 nm absorption results from V$_3$ centers forming in irradiated NaCl, which simultaneously strengthens the past identification of the 450 nm feature as an NaCl F-center absorption. 

The primary difference between the HST and laboratory spectra is the observed ratios of the F- and V$_3$-center bands. The laboratory ratio is nearly 1.8 times that observed anywhere within Tara Regio in the current data (6.9 in the lab vs. 3.9 where the strongest F-centers are observed). However, color-centers are dynamic and dependent on the instantaneous balance between destruction by solar photobleaching and creation via energetic particle bombardment. This balance is not well replicated in the lab and is time-variable on Europa \citep[e.g.][]{Paranicas2009, Bagenal2020}. As color-center absorptions depend on the radiation flux, and as the UV spectra of Tara Regio were taken two years after their visible-wavelength counterparts during likely different magnetospheric conditions, the HST ratio of the dynamic F- and V$_3$-center bands may be more misleading than meaningful. The data of \citet{Brown2022} also hint that differing effects of photobleaching on the F- and V$_3$- centers may shrink their ratio under true Europa conditions. In fact, concurrent laboratory work by \citet{Denman2022} investigating the dynamics of NaCl color centers in response to varying flux and photobleaching conditions shows that similar dynamics operating on timescales relevant to Europa's daily rotation likely also explain the lack of an M-center band in the visible-wavelength HST data, supporting the hypothesis posed at the time of the F-center detection \citep{TrumboEtAl2019}.

\section{Discussion} \label{sec:discussion}
The new detection of an NaCl V$_3$-center absorption, in combination with the parallel laboratory work explaining the HST data in the context of color-center formation and decay at Europa-like conditions \citep{Brown2022, Denman2022}, effectively confirms the presence of NaCl on the surface of Europa. In addition, the clear association of both the V$_3$- and F-center features with geologically disrupted chaos terrain implies that this NaCl originated in Europa's internal ocean, which has important implications for the nature of Europa's endogenous salts. NaCl within chaos terrain on the leading hemisphere is most consistent with the chloride hypothesis of \citet{BrownHand2013}, in which the endogenous salts emplaced onto the surface are chloride-rich, preserved as chloride-rich within the geology of the leading hemisphere, and converted radiolyticially to a more sulfate-rich composition within the chaos of the sulfur-bombarded trailing hemisphere. In contrast, the NaCl confirmation runs counter to the sulfate-rich interpretations of NIMS data \citep[e.g.][]{McCord1998_science, Dalton2012} and the magnesium-rich chlorate, perchlorate, and chloride mixtures of \citet{Ligier2016}, whose hypothesis places the highest abundances of chlorinated species on the trailing hemisphere, where we find no NaCl absorption. 

It is possible that the detectable presence of NaCl on the surface may be used to constrain the relative abundances of chloride (Cl$^-$) vs. sulfate (SO$_4^{2-}$) and sodium (Na$^+$) vs. magnesium (Mg$^{2+}$) in the source material \citep[e.g.][]{Johnson2019}, depending on the freezing history of the material and the mechanisms by which it was emplaced onto the surface. The brine-freezing experiments of \citet{Vu2016} suggest that, in the absence of flash freezing, Na$^+$, Mg$^{2+}$, Cl$^-$, and SO$_4^{2-}$ containing brines extruded onto the surface would only form abundant NaCl if the Na$^+$/Mg$^{2+}$ and Cl$^-$/SO$_4^{2-}$ ratios were high, as mirabilite (NaSO$_4$ $\cdot$ 10H$_2$O) and MgCl$_2$ were the primary phases frozen from their solutions. Alternatively, more rapid freezing rates \citep[e.g.][]{Vu2020} or the fractionation of upwelling ocean material \citep[e.g.][]{ZolotovShock2001} may complicate using such surface detections to place constraints on subsurface ionic abundances. However, regardless of how directly the NaCl relates to its relative abundance in the ocean, the concrete identification of endogenous chloride on Europa's surface warrants rethinking the long-standing idea that Europa's internal ocean may be sulfate-dominated \citep[e.g.][]{Kargel1991, Fanale2001}.

\section{Conclusions} \label{sec:conclusions}
Using HST/STIS spectroscopy, we detect a previously unseen absorption feature near 230 nm, which we attribute to V$_3$ centers of irradiated NaCl. The feature appears strongest within leading-hemisphere chaos terrain, particularly Tara Regio, and corresponds geographically to the proposed F-center absorption of NaCl detected in recent visible-wavelength HST data. The excellent spectral and spatial match between the 230 nm feature and NaCl confirms the presence of NaCl within geologically young terrain on Europa, implying the presence of NaCl in the interior ocean, and challenging past interpretations concerning the composition of Europa's endogenous salts.
\newline
\newline
\newline
\newline
\newline
\newline
\newline
\newline
\newline
\newline
\newline
\newline
\newline
\newline
\newline

\acknowledgments
Based on observations made with the NASA/ESA Hubble Space Telescope, obtained at the Space Telescope Science Institute, which is operated by the Association of Universities for Research in Astronomy, Inc., under NASA contract NAS5-26555. These observations are associated with program $\#$15095. This work is also associated with archival program $\#$15789. Support for programs $\#$15095 and $\#$15789 was provided by NASA through grants from the Space Telescope Science Institute, which is operated by the Association of Universities for Research in Astronomy, Inc., under NASA contract NAS5-26555. This work was also supported by grant number 668346 from the Simons Foundation. S.K.T. is supported by the Heising-Simons Foundation through a \textit{51 Pegasi b} postdoctoral fellowship. All of the data presented in this paper were obtained from the Mikulski Archive for Space Telescopes (MAST) at the Space Telescope Science Institute. The specific observations analyzed can be accessed via \dataset[10.17909/t9-5tsd-s772]{https://doi.org/10.17909/t9-5tsd-s772}.
\facility{HST(STIS)}
\software{astropy \citep{2013A&A...558A..33A}}

\end{document}